\documentclass[preprint,aps]{revtex4-2}
\usepackage{mathrsfs}
\usepackage{graphicx}
\usepackage{multirow}
\usepackage{makecell}
\usepackage{booktabs}
\usepackage{color}
\usepackage{bm}

\usepackage{hyperref}
\hypersetup{
    colorlinks,
    citecolor=black,
    filecolor=black,
    linkcolor=black,
    urlcolor=black
}
\begin{document}

\title{Disorder Driven Non-Anderson Transition in a Weyl Semimetal}

\author{Cong Li$^{1,*}$, Yang Wang$^{1}$, Jianfeng Zhang$^{2}$, Hongxiong Liu$^{2}$, Wanyu Chen$^{1}$, Guowei Liu$^{3}$, Hanbin Deng$^{3}$, Timur Kim$^{4}$, Craig Polley$^{5}$, Balasubramanian Thiagarajan$^{5}$, Jiaxin Yin$^{3}$, Youguo Shi$^{2}$, Tao Xiang$^{2}$, Oscar Tjernberg$^{1,*}$
}

\affiliation{
\\$^{1}$Department of Applied Physics, KTH Royal Institute of Technology, Stockholm 11419, Sweden
\\$^{2}$Beijing National Laboratory for Condensed Matter Physics, Institute of Physics, Chinese Academy of Sciences, Beijing 100190, China
\\$^{3}$Department of Physics, Southern University of Science and Technology, Shenzhen, Guangdong 518055, China
\\$^{4}$Diamond Light Source, Harwell Campus, Didcot, OX11 0DE, United Kingdom
\\$^{5}$MAX IV Laboratory, Lund University, 22100 Lund, Sweden
\\$^{*}$Corresponding authors: conli@kth.se, oscar@kth.se
}


\maketitle

%%Abstract

\begin{center}
{\bf Abstract}
\end{center}

{\bf For several decades, it was widely believed that a non-interacting disordered electronic system could only undergo an Anderson metal-insulator transition due to Anderson localization. However, numerous recent theoretical works have predicted the existence of a disorder-driven non-Anderson phase transition that differ from Anderson localization. The frustration lies in the fact that this non-Anderson disorder-driven transition has not yet been experimentally demonstrated in any system. Here, using angle-resolved photoemission spectroscopy, we present a case study of observing the non-Anderson disorder-driven transition by visualizing the electronic structure of the Weyl semimetal NdAlSi on surfaces with varying amounts of disorder. Our observations reveal that strong disorder can effectively suppress all surface states in the Weyl semimetal NdAlSi, including the topological surface Fermi arcs. This disappearance of surface Fermi arcs is associated with the vanishing of the bulk topological invariant, indicating a quantum phase transition from a Weyl semimetal to a diffusive metal. By analyzing the changes in the electronic structure of NdAlSi, as the surface degrades, we provide a physical picture of this non-Anderson transition from a Weyl semimetal to a diffuse metal. These observations provide the first direct experimental evidence of the non-Anderson disorder-driven transition, a discovery long anticipated by theoretical physicists. The finding dispels longstanding suspicions among researchers that non-Anderson transitions exist in real quantum systems.\\
}

%%Introduction
\noindent {\bf Introduction}

Disorder-driven transitions are a longstanding area of interest in the field of condensed matter physics. For a long time, it was generally believed that the Anderson localization\cite{PWAnderson_PR1958} transition is the only possible disorder-driven transition in non-interacting systems. Recently, however, extensive theoretical works have proposed that a broad class of systems may exhibit a new kind of disorder-driven transitions occurring prior to Anderson localization, which are manifested by the critical behaviour of the disorder-averaged density of states and other observable phenomena\cite{EFradkin_PRB1986_01,EFradkin_PRB1986_02,RShidou_PRB2009_SMurakami,PGoswami_PRL2011_SChakravarty,SRyu_PRB2012_KNomura,KKobayashi_PRL2014_IFHerbut,YOminato_PRB2014_MKoshino,BSbierski_PRL2014_PWBrouwer,JHPixley_PRL2015_SDSarma,BSbierski_PRB2015_PWBrouwer,SVSyzranov_PRL2015_VGurarie,SVSyzranov_PRB2015_LRadzihovsky,CZChen_PRL2015_XCXie,AAltland_PRL2015_DBagrets,JHPixley_PRB2016_SDSarma,SLiu_PRL2016_RShindou,HShapourian_PRB2016_TLHughes,SVSyzranov_PRB2016_LRadzihovsky,SBera_PRB2016_BRoy,JHPixley_PRX2016_SDSarma,SVSyzranov_AoP2016_LRadzihovsky,EVGorbar_PRB2016_POSukhachov,RJSlager_PRB2017_BRoy,MJPark_PRB2017_MJGilbert,JHWilson_PRB2018_SDSarma,BRoy_PRX2018_VJuricic,SVSyzranov_AR2018_LRzdzihovsky,MBuchhold_PRL2018_AAltland,MBuchhold_PRB2018_AAltland,NPArmitage_RMP2018_AVishwanath,JKlier_PRB2019_ADMirlin,CWang_PRR2020_XRWang,JYZhang_PRB2022_YLi}.  The theory suggests that such non-Anderson disorder-driven transition is a general phenomenon that occurs in all quantum systems in sufficiently high dimensions\cite{SVSyzranov_PRL2015_VGurarie,SVSyzranov_PRB2015_LRadzihovsky,SVSyzranov_AR2018_LRzdzihovsky}, attracting significant attention, especially in the study of three dimensional (3D) topological semimetals\cite{KKobayashi_PRL2014_IFHerbut,YOminato_PRB2014_MKoshino,BSbierski_PRL2014_PWBrouwer,JHPixley_PRL2015_SDSarma,BSbierski_PRB2015_PWBrouwer,SVSyzranov_PRL2015_VGurarie,SVSyzranov_PRB2015_LRadzihovsky,CZChen_PRL2015_XCXie,AAltland_PRL2015_DBagrets,JHPixley_PRB2016_SDSarma,SLiu_PRL2016_RShindou,HShapourian_PRB2016_TLHughes,SVSyzranov_PRB2016_LRadzihovsky,SBera_PRB2016_BRoy,JHPixley_PRX2016_SDSarma,SVSyzranov_AoP2016_LRadzihovsky,EVGorbar_PRB2016_POSukhachov,RJSlager_PRB2017_BRoy,MJPark_PRB2017_MJGilbert,JHWilson_PRB2018_SDSarma,BRoy_PRX2018_VJuricic,SVSyzranov_AR2018_LRzdzihovsky,MBuchhold_PRL2018_AAltland,MBuchhold_PRB2018_AAltland,NPArmitage_RMP2018_AVishwanath,JKlier_PRB2019_ADMirlin,CWang_PRR2020_XRWang,JYZhang_PRB2022_YLi}.

Topological semimetals are a class of materials that possess unique electronic properties due to their topological characteristics, including Dirac semimetals, Weyl semimetals, and nodal line semimetals\cite{NPArmitage_RMP2018_AVishwanath,KManna_NRM2018_CFelser,BQLv_RMP2021_HDing}. Recently, extensive theoretical studies have pointed out that Weyl semimetals are a potential platform for observing the non-Anderson disorder-driven transition\cite{KKobayashi_PRL2014_IFHerbut,YOminato_PRB2014_MKoshino,BSbierski_PRL2014_PWBrouwer,JHPixley_PRL2015_SDSarma,BSbierski_PRB2015_PWBrouwer,SVSyzranov_PRL2015_VGurarie,SVSyzranov_PRB2015_LRadzihovsky,CZChen_PRL2015_XCXie,AAltland_PRL2015_DBagrets,JHPixley_PRB2016_SDSarma,SLiu_PRL2016_RShindou,HShapourian_PRB2016_TLHughes,SVSyzranov_PRB2016_LRadzihovsky,SBera_PRB2016_BRoy,JHPixley_PRX2016_SDSarma,SVSyzranov_AoP2016_LRadzihovsky,EVGorbar_PRB2016_POSukhachov,RJSlager_PRB2017_BRoy,MJPark_PRB2017_MJGilbert,JHWilson_PRB2018_SDSarma,BRoy_PRX2018_VJuricic,SVSyzranov_AR2018_LRzdzihovsky,MBuchhold_PRL2018_AAltland,MBuchhold_PRB2018_AAltland,NPArmitage_RMP2018_AVishwanath,JKlier_PRB2019_ADMirlin,CWang_PRR2020_XRWang,JYZhang_PRB2022_YLi}. In Weyl semimetals, the topological surface Fermi arc (SFA) can survive weak disorder\cite{RJSlager_PRB2017_BRoy,JHWilson_PRB2018_SDSarma}. However, as the disorder increases beyond a certain threshold, the SFA will disappear\cite{RJSlager_PRB2017_BRoy,JHWilson_PRB2018_SDSarma}, accompanied by a non-Anderson disorder-driven quantum phase transition from the Weyl semimetal to a diffusive metal state\cite{CZChen_PRL2015_XCXie,HShapourian_PRB2016_TLHughes,RJSlager_PRB2017_BRoy,BRoy_PRX2018_VJuricic,JYZhang_PRB2022_YLi}. However, so far, no experiments have observed the non-Anderson transition where disorder causes the disappearance of SFA in Weyl semimetals. On the contrary, experiments have demonstrated that SFA remains robust even in the presence of disorder\cite{PSessi_PRB2017_MBode}. In fact, over the past decade, the existence of non-Anderson disorder-driven quantum phase transitions also has been predicted in various quantum systems beyond topological semimetals, including 1D and 2D arrays of ultracold trapped ions, chiral superconductors and quantum kicked rotors, and numerous new theoretical predictions continue to emerge\cite{EFradkin_PRB1986_01,EFradkin_PRB1986_02,RShidou_PRB2009_SMurakami,PGoswami_PRL2011_SChakravarty,SRyu_PRB2012_KNomura,KKobayashi_PRL2014_IFHerbut,YOminato_PRB2014_MKoshino,BSbierski_PRL2014_PWBrouwer,JHPixley_PRL2015_SDSarma,BSbierski_PRB2015_PWBrouwer,SVSyzranov_PRL2015_VGurarie,SVSyzranov_PRB2015_LRadzihovsky,CZChen_PRL2015_XCXie,AAltland_PRL2015_DBagrets,JHPixley_PRB2016_SDSarma,SLiu_PRL2016_RShindou,HShapourian_PRB2016_TLHughes,SVSyzranov_PRB2016_LRadzihovsky,SBera_PRB2016_BRoy,JHPixley_PRX2016_SDSarma,SVSyzranov_AoP2016_LRadzihovsky,EVGorbar_PRB2016_POSukhachov,RJSlager_PRB2017_BRoy,MJPark_PRB2017_MJGilbert,JHWilson_PRB2018_SDSarma,BRoy_PRX2018_VJuricic,SVSyzranov_AR2018_LRzdzihovsky,MBuchhold_PRL2018_AAltland,MBuchhold_PRB2018_AAltland,NPArmitage_RMP2018_AVishwanath,JKlier_PRB2019_ADMirlin,CWang_PRR2020_XRWang,JYZhang_PRB2022_YLi}. However, experimental evidence for the non-Anderson transition has been frustratingly absent in any quantum system to date. This has led to suspicions that the non-Anderson disorder-driven transition might be nothing more than a theoretical conjecture or purely mathematical model conceived by theoretical physicists.

Here, we take advantage of the surface-sensitive nature of angle-resolved photoemission spectroscopy (ARPES) and observe direct evidence of a non-Anderson disorder-driven quantum phase transition in a quasi-two-dimensional system near the surface. Employing ARPES and scanning tunneling microscope (STM) measurements as well as density functional theory (DFT) calculations, we systematically investigate the near surface electronic structure of the magnetic Weyl semimetal NdAlSi\cite{JGaudet_NM2021_CLBroholm,CLi_NC2023_OTjernberg,CLi_arxiv2024_OTjernberg}, on surfaces with varying amount of disorder. Our findings reveal that all surface states, including the topological surface Fermi arcs (SFAs), are completely suppressed on the strong disordered surface of NdAlSi. To experimentally simulate increasing disorder, we conducted time-dependent electronic structure measurements on a highly ordered surface of NdAlSi. The results demonstrate that the SFAs can survive weak disorder, but the shape of the SFAs change. In strong disorder, the SFAs completely dissolve into the bulk metallic background, indicating the emergence of the non-Anderson disorder-driven quantum phase transition from a Weyl semimetal to a diffusive metal in a quasi-two-dimensional system near the surface. The transition is induced by gradually increasing the disorder of the top-most atomic layers of the system through adsorption or surface destruction. Our findings provide first direct experimental evidence for the existence of the non-Anderson disorder-driven transition in quantum systems, transcending purely mathematical concepts.\\

%Figure1

\begin{figure*}[tbp]
\begin{center}
\includegraphics[width=0.85\columnwidth,angle=0]{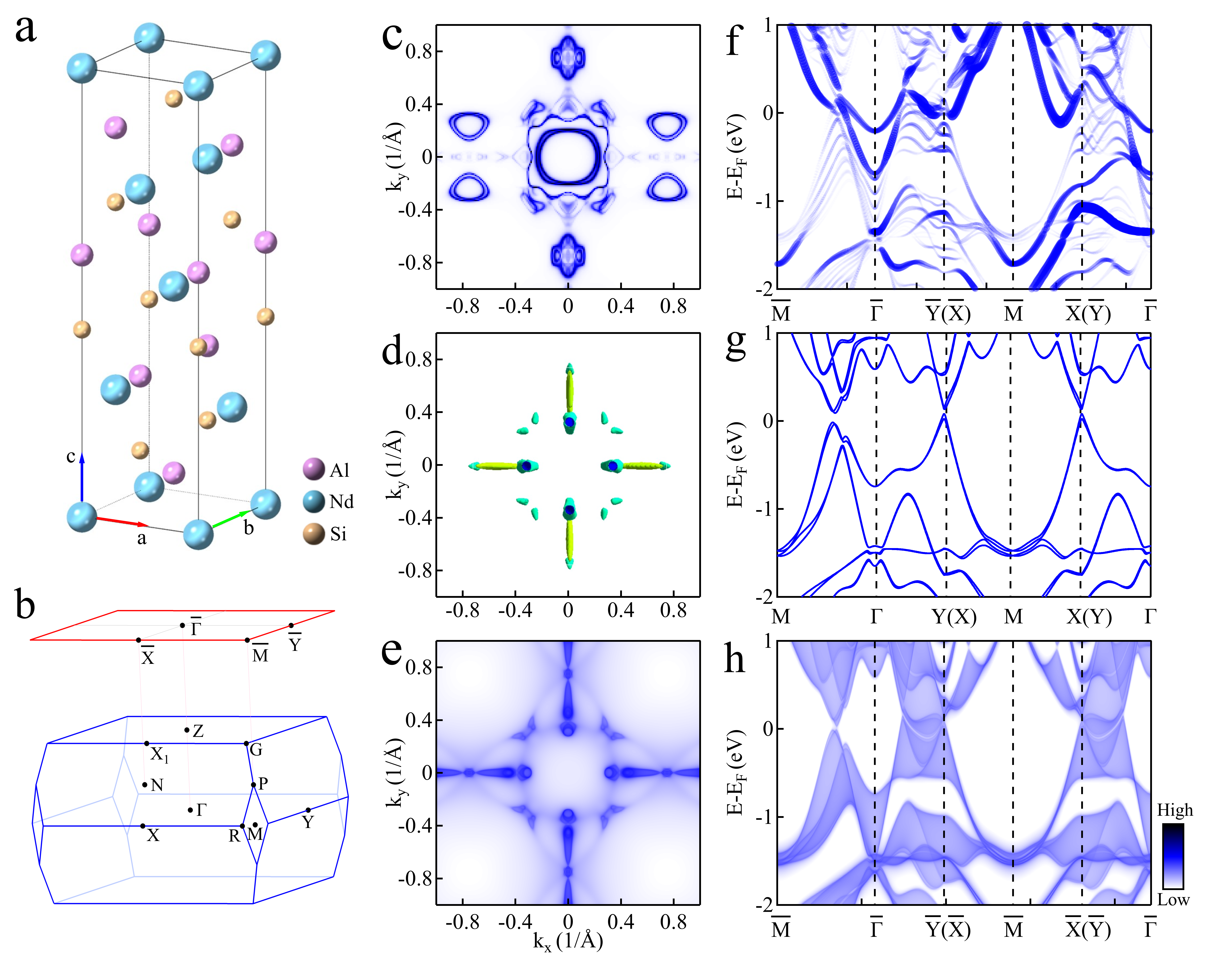}
\end{center}
\caption{\footnotesize\textbf{Crystal structure and calculated electronic structure of NdAlSi.} (a) The crystal structure of NdAlSi with the space group $I4_{1}md$ (no. 109). (b) The 3D BZ of the original unit cell of NdAlSi, and the corresponding two-dimensional BZ projected on the (001) plane (red lines) in the pristine phase in (a). (c) Surface projected DFT calculated Fermi surface on the terminal surface of Nd atoms cleavage at the Al-Nd layer. (d) The DFT calculated 3D bulk Fermi surface of NdAlSi. (e) The DFT calculated bulk Fermi surface of NdAlSi which integrate all of the BZ along k$_z$ direction. (f) The surface projected DFT calculated band dispersion along $\overline{M}-\overline{\Gamma}-\overline{Y}-\overline{M}-\overline{X}-\overline{\Gamma}$ directions on the terminal surface of Nd atoms cleavage at the Al-Nd layer. (g) Calculated band structures of NdAlSi along high-symmetry directions across the BZ. (h) The DFT calculated bulk band structure of NdAlSi which integrate all of the BZ along k$_z$ direction.
}
\label{1}
\end{figure*}

To achieve a comprehensive understanding of the electronic structure of NdAlSi, we first performed DFT calculations on it, as shown in Fig.~\ref{1}. Fig.~\ref{1}a shows the crystal structure of NdAlSi, which crystallizes in the tetragonal structure with the space group I4$_1$md (no. 109)\cite{JGaudet_NM2021_CLBroholm}. The corresponding three-dimensional (3D) Brillouin zone (BZ) of NdAlSi is shown in Fig.~\ref{1}b, which defines the high symmetry points in the BZ. According to previous reports, the Al-Nd layer is the dominating cleavage plane in NdAlSi\cite{CLi_NC2023_OTjernberg,CLi_arxiv2024_OTjernberg}. DFT calculations predict that the Nd terminated surface from a cleave at the Al-Nd layer forms a complex surface state (Fig.~\ref{1}c and~\ref{1}f) which differs from the electronic structure of the bulk state (Fig.~\ref{1}d and~\ref{1}g)\cite{CLi_NC2023_OTjernberg}. Furthermore, DFT calculations of the bulk states, integrated along the k$_z$ direction as shown in Fig.~\ref{1}e and~\ref{1}h, provide an overview of the bulk and surface electronic structure of NdAlSi.

%Figure2

\begin{figure*}[tbp]
\begin{center}
\includegraphics[width=0.85\columnwidth,angle=0]{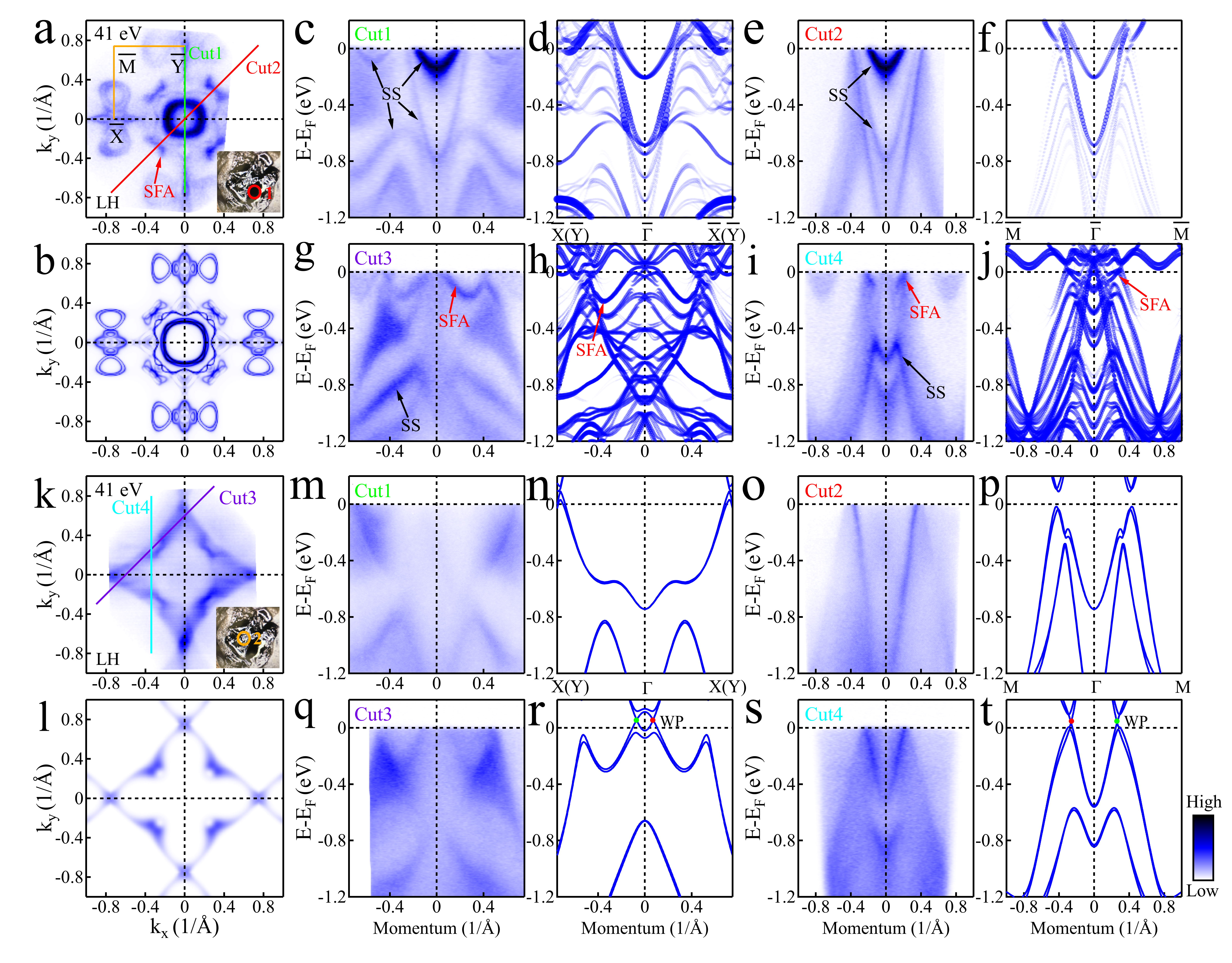}
\end{center}
\caption{\footnotesize\textbf{Surface and bulk electronic structures of NdAlSi.} (a) Fermi surface of NdAlSi measured with photon energy of 41 eV under LH polarization in the area 1 (red circle) of sample. (b) Surface projected DFT calculated Fermi surface on the terminal surface of Nd atoms cleavage at the Al-Nd layer with considering two domain structures. (c), (e), (g) and (i) Measured band dispersions along $\overline{Y}-\overline{\Gamma}-\overline{Y}$ [Cut1, (c)], $\overline{M}-\overline{\Gamma}-\overline{M}$ [Cut2, (e)], Cut3 and Cut4 directions in the area 1 of sample under LH polarization. Cut3 and Cut4 go right through the Weyl points. (d), (f), (h) and (j) The corresponding surface projected DFT calculations of the bands along $\overline{Y}-\overline{\Gamma}-\overline{Y}$ (d), $\overline{M}-\overline{\Gamma}-\overline{M}$ (f), Cut3 (h) and Cut4 (j) directions, considering the two domain structures. (k) Fermi surface of NdAlSi measured with photon energy of 41 eV under LH polarization in the area 2 (orange circle) of sample. According to the previous report, bulk state measurements with photon energy of 41 eV corresponds to the k$_z$ $\sim$ 0 $\pi$/c plane\cite{CLi_NC2023_OTjernberg}. (l) The DFT calculated bulk Fermi surface at the k$_z$= 0 $\pi/c$ plane which integrate 0$\pm$0.1 $\pi/c$ of BZ along k$_z$ direction. (m), (o), (q) and (s) The similar measurements as (c), (e), (g) and (i) but measured in the area 2 of sample. (n), (p), (r) and (t) The DFT calculated bulk band structures along $Y-\Gamma-Y$ (n), $M-\Gamma-M$ (p), Cut3 (r) and Cut4 (t) directions. The trivial surface states (SSs) are marked by black arrows, and the SFAs are marked by red arrows which was well studied in Ref.\cite{CLi_NC2023_OTjernberg}. The green and red dots in (r) and (t) mark the Weyl points (WPs), while the red dots representing nodes with chiralities +1 and green dots representing -1.
}
\label{2}
\end{figure*}

In conjunction with DFT calculations, we performed ARPES measurements on NdAlSi. Since NdAlSi is a 3D non-centrosymmetric crystal, the single crystal after cleavage shows flat areas (area 1 in Fig.~\ref{2}a) as well as areas with multiple steps (area 2 in Fig.~\ref{2}k). The corresponding location can be found by spatial scanning of the sample. When the measurement position is in area 1 (red circle in Fig.~\ref{2}a), the measured Fermi surface is as shown in Fig.~\ref{2}a. It is seen that the measured Fermi surface (Fig.~\ref{2}a) is in good agreement with the surface projected DFT Fermi surface calculations for the Nd terminated surface at the Al-Nd layer, taking into account the two domain structures (Fig.~\ref{2}b) visible due to the fact that the light spot covers two orthogonal domains simultaneously\cite{CLi_NC2023_OTjernberg}. When the two orthogonal domain structures leading to superposition of bands along the $\overline{X}-\overline{\Gamma}-\overline{X}$ and $\overline{Y}-\overline{\Gamma}-\overline{Y}$ directions are taken into account, the measured band dispersions along $\overline{Y}-\overline{\Gamma}-\overline{Y}$ [Cut1 (green line in Fig.~\ref{2}a), Fig.~\ref{2}c], $\overline{M}-\overline{\Gamma}-\overline{M}$ [Cut2 (red line in Fig.~\ref{2}a), Fig.~\ref{2}e], Cut3 (purple line in Fig.~\ref{2}k, Fig.~\ref{2}g) and Cut4 (cyan line in Fig.~\ref{2}k, Fig.~\ref{2}i) directions can be mapped onto the corresponding surface projected DFT calculated band structure (Fig.~\ref{2}d, Fig.~\ref{2}f, Fig.~\ref{2}h and Fig.~\ref{2}j)\cite{CLi_NC2023_OTjernberg}. However, when the measurement position is in area 2 (orange circle in Fig.~\ref{2}k), the measured Fermi surface appears as in Fig.~\ref{2}k. It is strikingly different from the measured Fermi surface in area 1 (Fig.~\ref{2}a). In order to further understand the reasons behind the difference between these two measured Fermi surfaces (Fig.~\ref{2}a and~\ref{2}k), measurements corresponding to those in Fig.~\ref{2}c,~\ref{2}e,~\ref{2}g and~\ref{2}i were taken in area 2 (orange circle in Fig.~\ref{2}f) and shown in Fig.~\ref{2}m,~\ref{2}o,~\ref{2}q and~\ref{2}s. We note that some band features in Fig.~\ref{2}c,~\ref{2}e,~\ref{2}g and~\ref{2}i (marked by black and red arrows) are lacking in Fig.~\ref{2}m,~\ref{2}o,~\ref{2}q and~\ref{2}s. To clarify this, DFT bulk band structure calculations were carried out along $X(Y)-\Gamma-X(Y)$ (Fig.~\ref{2}n), $M-\Gamma-M$ (Fig.~\ref{2}p), Cut3 (Fig.~\ref{2}r) and Cut4 (Fig.~\ref{2}t). As a result, we find that all the band features in the measured band dispersions (Fig.~\ref{2}m,~\ref{2}o,~\ref{2}q and~\ref{2}s) are captured by the corresponding bulk band structure calculations (Fig.~\ref{2}n,~\ref{2}p,~\ref{2}r and~\ref{2}t). Furthermore, we performed DFT calculations for the bulk Fermi surface at the k$_z$= 0 $\pi/c$ plane, integrating 0$\pm$0.1 $\pi/c$ of the BZ along the k$_z$ direction, as shown in Fig.~\ref{2}l. The calculated bulk Fermi surface (Fig.~\ref{2}l) is also in good agreement with the Fermi surface measured in area 2 with a photon energy of 41 eV (Fig.~\ref{2}k).

It appears clear that the electronic structure measured in area 1 (the flat area, Fig.~\ref{2}a,~\ref{2}c, ~\ref{2}e,~\ref{2}g and~\ref{2}i) is mainly from the surface states and the electronic structure measured in area 2 (the area with multiple steps, Fig.~\ref{2}k,~\ref{2}m,~\ref{2}o,~\ref{2}q and~\ref{2}s) is mainly from the bulk states. This provides us with an opportunity to directly study the bulk electronic structure of NdAlSi with high energy and momentum resolution. Furthermore, it is worth noting that in addition to the trivial surface states, the SFAs measured on the ordered surface also completely disappear on the disordered surface (Fig.~\ref{2}q and~\ref{2}s). In addition, the bulk states measured on the disordered surface do not show any obvious degradation or broadening as compared with the flat surface, some bands even look sharper. A more detailed analysis is given in the Supplementary Material (Fig.~S1). We note  that in the context of a three dimensional (3D) Weyl semimetal, the bulk boundary correspondence highlights that Weyl points are always associated with SFAs. The disappearance of SFAs connecting the projected Weyl points in area 2 seem to challenge the bulk boundary correspondence.

%Figure3

\begin{figure*}[tbp]
\begin{center}
\includegraphics[width=1\columnwidth,angle=0]{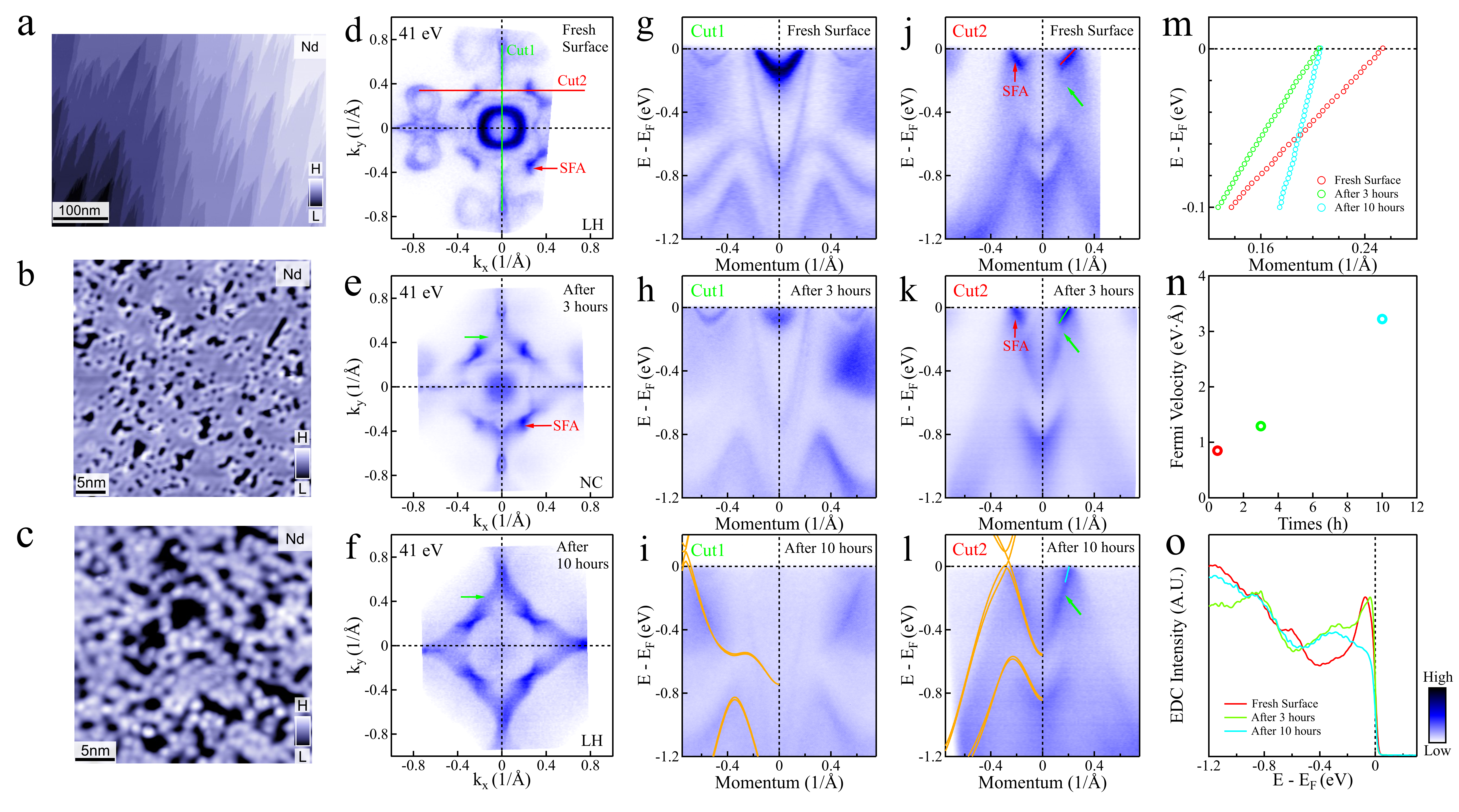}
\end{center}
\caption{\footnotesize\textbf{The STM measurements and evolution of the electronic structure with time.} (a) STM images of NdAlSi measured at 4.7 K on Nd atom terminated surface cleavage at the Al-Nd layer. (a) is the overview image. (b) and (c) are the magnified topography images measured at different positions. Scan conditions: (a) 1 V, 0.1 nA, (b) 0.5 V, 0.5 nA, (c) 0.01V, 0.1 nA. (d-g) Fermi surface of NdAlSi measured with photon energy of 41 eV under LH (d,f,g) and NC (e) polarizations on flat area of the fresh sample (d), the sample 3 hours after cleavage (e), the sample 10 hours after cleavage (f). (g-i) The corresponding band dispersions along $\overline{Y}-\overline{\Gamma}-\overline{Y}$ [Cut1, green line in (d)] directions. (j-l) The corresponding bands along Cut2 [red line in (d)] directions. The corresponding DFT bulk calculations are overlaid on (i) and (l). The red arrows mark the SFA. (m) Band dispersions extracted from (j-l) by fitting the energy dependent MDCs. The corresponding fitting band dispersions are also appended to (j-l). (n) The time dependent Fermi velocity of SFAs which obtained by linear fitting the band dispersions in (m). (o) The normalized integrated EDC of bands in (j-l) at energy range of -0.9 eV to -0.7 eV.
}
\label{3}
\end{figure*}

In order to better understand the mechanism behind this, we carried out STM measurements as shown in Fig.~\ref{3}a-\ref{3}c. Fig.~\ref{3}a shows the overview STM image of the Nd terminated surface which exhibit multiple steps. In this area, we choose two different positions to do magnified topography image scans, as shown in Fig.~\ref{3}b and~\ref{3}c. Considerable disorder is observed in the form of atomic clusters and holes. In theoretical simulations\cite{RJSlager_PRB2017_BRoy,JHWilson_PRB2018_SDSarma}, it has been found that although the SFAs in Weyl semimetals are topologically protected, they are not robust against strong disorder. While the SFAs can survive in weak disorder, they dissolve into the bulk metallic background as the disorder increases\cite{RJSlager_PRB2017_BRoy,JHWilson_PRB2018_SDSarma}. As the SFAs dissolve, the density of state (DOS) of low-energy electron states near the Fermi energy will gradually decrease. In strong disorder, the disappearance of the SFAs is directly related to the vanishing of the bulk topological invariant. Therefore, the disappearance of the SFA is accompanied by a Weyl semimetal-diffusive metal quantum phase transition. The metallic phase is also supporting quasi-particles with a finite lifetime and mean free path, but is topologically trivial. Thus, the disappearance of the SFAs does not invalidate the bulk-boundary correspondence\cite{RJSlager_PRB2017_BRoy}. 

To study the electronic structure evolution with increasing disorder, we carried out time dependent electronic structure measurements in the ordered areas of the sample. In this case, all of the surface states (trivial surface states and SFAs) can be clearly observed when measured on the ordered and freshly cleaved sample, as shown in Fig.~\ref{3}d,~\ref{3}g and~\ref{3}j. However, three hours later, the surface states appear blurred (Fig.~\ref{3}e,~\ref{3}h and~\ref{3}k) relative to the newly cleaved sample, and some structures of the surface state (the electron like band in the center of the BZ, Fig.~\ref{3}h) even shift to higher binding energy. Ten hours later, all of the surface states (trivial surface states and SFAs) seem to be completely suppressed, leaving only the bulk states, as shown in Fig.~\ref{3}f,~\ref{3}i and~\ref{3}l. This is similar to the electronic structure measured on the disordered surface of the newly cleaved sample (Fig.~\ref{2}k,~\ref{2}m and~\ref{2}s). The bulk states, on the other hand show no significant degradation relative to that measured on the fresh surface, and some band features even look sharper (marked by green arrows in Fig.~\ref{3}j-\ref{3}l), for more details see Fig. S2 in Supplementary Material. A reasonable explanation for the evolution of the above electronic structure over time is that the impurities are adsorbed at low temperatures, which would remove charge and weaken the bonds between Nd atoms and the surface. Consequently, this would increase the mobility of Nd atoms, leading to the formation of irregular clusters on the surface. This leads to surface disorder on the sample, thus affecting the near surface electronic structure of the sample. During the same time, the SFAs also evolve. In the beginning, the SFAs appear very sharp (Fig.~\ref{3}j), and after 3 hours they are gradually smoothed out (Fig.~\ref{3}k) until they disappeared completely after 10 hours (Fig.~\ref{3}l). In order to observe the changes of the SFAs more quantitatively, the dispersion of SFAs is obtained by fitting the energy-dependent momentum distribution curves (MDCs) of the SFAs, as shown in Fig.~\ref{3}m. Fig.~\ref{3}n shows the Fermi velocity of the SFAs and bulk bands. As the disorder increases, the Fermi velocity of the SFAs gradually increases until they completely dissolve with the bulk band. At the same time, the diamond-shaped Fermi surface (marked by green arrows in Fig.~\ref{3}e-\ref{3}f) gradually appears and eventually dominate as shown in Fig.~\ref{3}f. Fig.~\ref{3}o shows the integrated energy distribution curves (EDCs) of the bands in Fig.~\ref{3}j-\ref{3}l. It demonstrates that the near Fermi energy DOS gradually decreases with increasing disorder. 

%Figure4

\begin{figure*}[tbp]
\begin{center}
\includegraphics[width=1\columnwidth,angle=0]{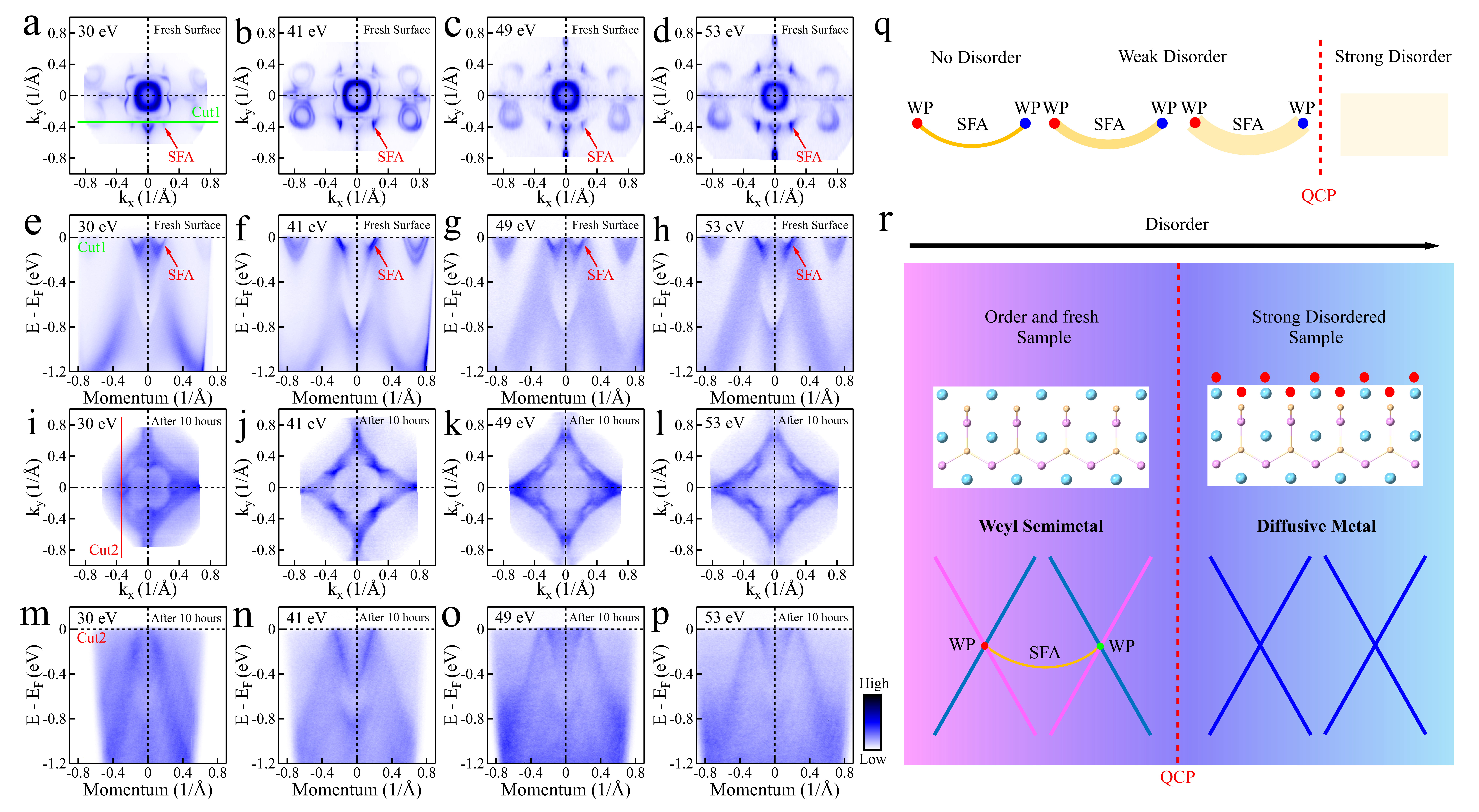}
\end{center}
\caption{\footnotesize\textbf{Weyl semimetal-diffusive metal quantum phase transition.} (a-d) Photon energy dependent Fermi surface measured on fresh surface of flat sample with photon energies of 30 eV (a), 41 eV (b), 49 eV (c) and 53 eV (d). (e-h) Photon energy dependent band
structures measured on fresh surface of flat sample with photon energies of 30 eV (e), 41 eV (f), 49 eV (g) and 53 eV (h) along Cut1 in (a). The red arrows mark the SFAs. (i-p) The similar measurements as (a-k) but measured on the surface 10 hours after the sample cleaved. (q) Schematic diagram of SFA suppressed by disorder. (r) Schematic phase diagram of the Weyl semimetal as the increase of disorder. The red dashed lines mark the quantum critical point (QCP).
}
\label{4}
\end{figure*}

In order to further study and disentangle the bulk vs surface electronic structure in NdAlSi, we performed photon energy-dependent Fermi surface measurements, as shown in Fig.~\ref{4}a-\ref{4}p. Fig.~\ref{4}a-\ref{4}d show the photon energy dependent Fermi surfaces measured on the fresh and ordered surface with photon energies of 30 eV (Fig.~\ref{4}a), 41 eV (Fig.~\ref{4}b), 49 eV (Fig.~\ref{4}c) and 53 eV (Fig.~\ref{4}d). The corresponding photon energy dependent band structures along Cut1 (green line in Fig.~\ref{4}a) are shown in Fig.~\ref{4}e-\ref{4}f. The SFAs are clearly observed at all photon energies (marked by red arrows in Fig.~\ref{4}a-\ref{4}h). After 10 hours, as the disorder increases, the SFAs completely dissolves into the bulk metallic background. At this point the SFAs completely disappear from the photon energy-dependent measurements (Fig.~\ref{4}i-\ref{4}p). The process of SFA dissolution is schematically shown in Fig.~\ref{4}q\cite{RJSlager_PRB2017_BRoy}. At first, the SFAs show very sharp features (Fig.~\ref{3}d, ~\ref{3}j,~\ref{4}a-\ref{4}h). In weak disorder, the SFAs still survive (Fig.~\ref{3}e,~\ref{3}k), but become less well defined as the disorder increases. When the disorder increases sufficiently, the SFAs completely disappear. At the same time, the diamond-shaped Fermi surface appears (Fig.~\ref{4}i-\ref{4}l). According to the previous report\cite{CLi_NC2023_OTjernberg}, bulk state measurements with photon energy of 41 eV corresponds to k$_z$ $\sim$ 0 $\pi$/c plane, the photon energy of 30 eV and 53 eV corresponds to kz $\sim$ 1 $\pi$/c plane and the photon energy of 49 eV corresponds to k$_z$ $\sim$ 0.67 $\pi$/c. The observed diamond-shaped Fermi surface is difficult to attribute to the bulk state dispersion, since it exhibit negligible photon energy dependence. 

Based on the above observations, we propose a conjecture for the underlying process. At first, on the clean ordered surface, the electronic structure is that of a Weyl semimetal. As the disorder increases at the surface, multiple boundaries are created that carry chiral edge states. For weak disorder, a larger number of chiral edge states can be accommodated that smooths out the SFAs. When the surface disorder increases sufficiently, the underlying second ordered layer becomes the topmost ordered layer. This layer does not have full direct contact with the vacuum and lacks the out-of-plane translational symmetry of the previous bulk state. Consequently, the associated two-dimensional electronic structure associated with this layer, characterized by the diamond-shaped Fermi surface, retains some aspects of the electronic structure of the bulk state. This view points to the measured electronic structures (Fig.~\ref{4}i-\ref{4}p) originating from the underlying second layer. At this point, the topological invariants of the underlying second ordered layer disappear, meaning that it no longer has the properties of a Weyl semimetal. As adsorption increases, the underlying bulk termination layer becomes further distanced from the vacuum.

According to the above analysis, the Weyl semimetal-diffusive metal quantum phase transition observed here occurs only in a few atomic layers on the surface. The main reason is that as the number of atomic layers increases, the disorder caused adsorption will decrease, so it can only affect a few atomic layers on the surface at most. It is due to the surface sensitivity of ARPES that we observe the non-Anderson disorder-driven Weyl semimetal-diffusive metal quantum phase transition in a quasi-two-dimensional system, gradually introducing enough disorder in the quasi-two-dimensional system of a few unit cells below the surface by adsorption or destruction of the surface. To the best of our knowledge, this non-Anderson disorder-driven transition has so far not been observed experimentally in any quantum system, despite considerable work on theoretical and numerical simulations\cite{EFradkin_PRB1986_01,EFradkin_PRB1986_02,RShidou_PRB2009_SMurakami,PGoswami_PRL2011_SChakravarty,SRyu_PRB2012_KNomura,KKobayashi_PRL2014_IFHerbut,YOminato_PRB2014_MKoshino,BSbierski_PRL2014_PWBrouwer,JHPixley_PRL2015_SDSarma,BSbierski_PRB2015_PWBrouwer,SVSyzranov_PRL2015_VGurarie,SVSyzranov_PRB2015_LRadzihovsky,CZChen_PRL2015_XCXie,AAltland_PRL2015_DBagrets,JHPixley_PRB2016_SDSarma,SLiu_PRL2016_RShindou,HShapourian_PRB2016_TLHughes,SVSyzranov_PRB2016_LRadzihovsky,SBera_PRB2016_BRoy,JHPixley_PRX2016_SDSarma,SVSyzranov_AoP2016_LRadzihovsky,EVGorbar_PRB2016_POSukhachov,RJSlager_PRB2017_BRoy,MJPark_PRB2017_MJGilbert,JHWilson_PRB2018_SDSarma,BRoy_PRX2018_VJuricic,SVSyzranov_AR2018_LRzdzihovsky,MBuchhold_PRL2018_AAltland,MBuchhold_PRB2018_AAltland,NPArmitage_RMP2018_AVishwanath,JKlier_PRB2019_ADMirlin,CWang_PRR2020_XRWang,JYZhang_PRB2022_YLi}. 
Our findings provide an experimental way to observe and study such non-Anderson disorder-driven transitions, which can provide new insights into quantum phase transitions driven by disorder. This certainly increases the confidence of researchers that this non-Anderson disorder transition is real in the actual material and not a purely mathematical concept. 

%Summary

In summary, by carrying out ARPES measurements on the Weyl semimetal NdAlSi, we found that all of the surface states, trivial surface states as well as SFAs, were suppressed on strongly disordered surfaces. The temporal evolution of the electronic structure observed on originally clean and ordered surfaces provides an experimental way of following increasing disorder in this Weyl semimetal. It was found that the SFAs can survive in weak disorder but as the disorder increases, the SFAs dissolve into the bulk metallic background until they disappear completely. The disappearance of the SFAs are related to the vanishing of the bulk topological invariant, indicating a Weyl semimetal-diffusive metal quantum phase transition. Our results provide direct evidence of non-Anderson disorder-driven transitions in Weyl semimetal NdAlSi, demonstrating that this transition is no longer merely a theoretical model. This discovery also opens new avenues for the study of new quantum states in topological materials caused by non-Andersion transitions.\\

%Methods

\noindent {\bf Methods}\\
\noindent{\bf Sample} Single crystals of NdAlSi were grown from Al as flux. Nd, Al, Si elements were sealed in an alumina crucible with the molar ratio of 1 : 10 : 1. The crucible was finally sealed in a highly evacuated quartz tube. The tube was heated up to 1273 K, maintained for 12 hours and then cooled down to 973 K at a rate of 3 K per hour. Single crystals were separated from the flux by centrifuging. The Al flux attached to the single crystals were removed by dilute NaOH solution.

\noindent{\bf ARPES Measurements} High-resolution ARPES measurements were performed at the I05 beamline of the Diamond synchrotron and at the Bloch beamline of MAX IV synchrotron. The total energy resolution (analyzer and beamline) was set at 15$\sim$20 meV for the measurements. The angular resolution of the analyser was $\sim$0.1 degree. The beamline spot size on the sample was about 70 $\mu$m$\times$70 $\mu$m at the I05 beamline of the Diamond synchrotron and about 10 $\mu$m$\times$12 $\mu$m at the Bloch beamline of the MAX IV synchrotron. The samples were cleaved {\it in situ} and measured in ultrahigh vacuum with a base pressure better than 1.0$\times$10$^{-10}$ mbar at about 10 K at the I05 beamline of the Diamond synchrotron and about 18 K at the Bloch beamline of the MAX IV synchrotron.

\noindent{\bf STM Measurements} STM experiments were conducted using an ultrahigh vacuum (UHV) low-temperature STM (Unisoku, USM1300) under a base pressure below 1$\times$10$^{-10}$ mbar. High-quality NdAlSi single crystals, measuring up to 3 mm$\times$3 mm$\times$3 mm, were bisected prior to being affixed to beryllium copper sheets. The crystal was mechanically cleaved in situ at 78 K and promptly placed into the microscope head, which was already at the base temperature of He4 (4.7 K). Topographic images were obtained using Ir/Pt tips in constant-current mode at 4.7 K.

\noindent{\bf DFT calculations} The electronic structure calculations for NdAlSi were performed based on the density functional theory (DFT)\cite{PHohenberg_PR1964_WKohn,WKohn_PR1965_LJSham} as implemented in the VASP package\cite{GKresse_CMS1996_JFurthmuller,GKresse_PRB1996_JFurthmuller}. The generalized gradient approximation (GGA) of Perdew-Burke-Ernzerhof (PBE) type\cite{JPPerdew_PRL1996_MErnzerhof} was chosen for the exchange-correlation functional. The projector augmented wave (PAW) method\cite{PEBlochl_PRB1994,GKresse_PRB1998_DJoubert} was adopted to describe the interactions between valence electrons and nuclei. In calculation, the Nd pseudopotential was chosen without the 4$f$ electrons. The kinetic energy cutoff of the plane-wave basis was set to be 350 eV. A 16$\times$16$\times$16 Monkhorst-Pack grids\cite{HJMonkhorst_PRB1976_JDPack} was used for the BZ sampling. For describing the Fermi-Dirac distribution function, a Gaussian smearing of 0.05 eV was used. When studying the surface states of NdAlSi, we employed a 36 atomic layers slab system with 20 \AA\ vacuum layer.

\noindent {\bf Data Availability}

\noindent The authors declare that all data supporting the findings of this study are available within the paper and its Supplementary Information files.

\vspace{3mm}

\noindent {\bf Acknowledgement}\\
The work presented here was financially supported by the Swedish Research council (2019-00701) and the Knut and Alice Wallenberg foundation (2018.0104). Y.G.S. acknowledges the National Natural Science Foundation of China (Grants No. U2032204), and the Informatization Plan of Chinese Academy of Sciences (CAS-WX2021SF-0102). We acknowledge MAX IV Laboratory for time on Beamline BLOCH under Proposal 20221199, 20230262 and 20231119. Research conducted at MAX IV, a Swedish national user facility, is supported by the Swedish Research council under contract 2018-07152, the Swedish Governmental Agency for Innovation Systems under contract 2018-04969, and Formas under contract 2019-02496.

\vspace{3mm}

\noindent {\bf Author Contributions}\\
C.L. proposed and conceived the project. C.L. carried out the ARPES experiments with the assistance from Y.W. and W.Y.C.. J.F.Z. and T.X. contributed to the band structure calculations. H.X.L. and Y.G.S. contributed to NdAlSi crystal growth. G.W.L., H.B.D. and J.X.Y. carried out the STM experiments. C.L. contributed to software development for data analysis and analyzed the data with Y.W.. C.L. wrote the paper. T.K., C.P. and B.T. provided the beamline support. Y.W. and O.T. participate in the scientific discussions. O.T. revised the manuscript. All authors participated in and commented on the paper.

\noindent {\bf Competing Interests}\\
The authors declare no competing interests.

\newpage

\end{document}

\centerline{\large{Supplementary Material for}}
%\begin{center}
%\Large Supplementary Material for
%\end{center}

\title{Disorder Driven Non-Anderson Transition in a Weyl Semimetal}

\author{Cong Li$^{1,*}$, Yang Wang$^{1}$, Jianfeng Zhang$^{2}$, Hongxiong Liu$^{2}$, Wanyu Chen$^{1}$, Guowei Liu$^{3}$, Hanbin Deng$^{3}$, Timur Kim$^{4}$, Craig Polley$^{5}$, Balasubramanian Thiagarajan$^{5}$, Jiaxin Yin$^{3}$, Youguo Shi$^{2}$, Tao Xiang$^{2}$, Oscar Tjernberg$^{1,*}$
}

\affiliation{
\\$^{1}$Department of Applied Physics, KTH Royal Institute of Technology, Stockholm 11419, Sweden
\\$^{2}$Beijing National Laboratory for Condensed Matter Physics, Institute of Physics, Chinese Academy of Sciences, Beijing 100190, China
\\$^{3}$Department of Physics, Southern University of Science and Technology, Shenzhen, Guangdong 518055, China
\\$^{4}$Diamond Light Source, Harwell Campus, Didcot, OX11 0DE, United Kingdom
\\$^{5}$MAX IV Laboratory, Lund University, 22100 Lund, Sweden
\\$^{*}$Corresponding authors: conli@kth.se, oscar@kth.se
}

\pacs{}

\maketitle

{\bf 1. Comparison of constant energy contours and band structures of flat and uneven surfaces}

\begin{figure*}[tbp]
\begin{center}
\includegraphics[width=1\columnwidth,angle=0]{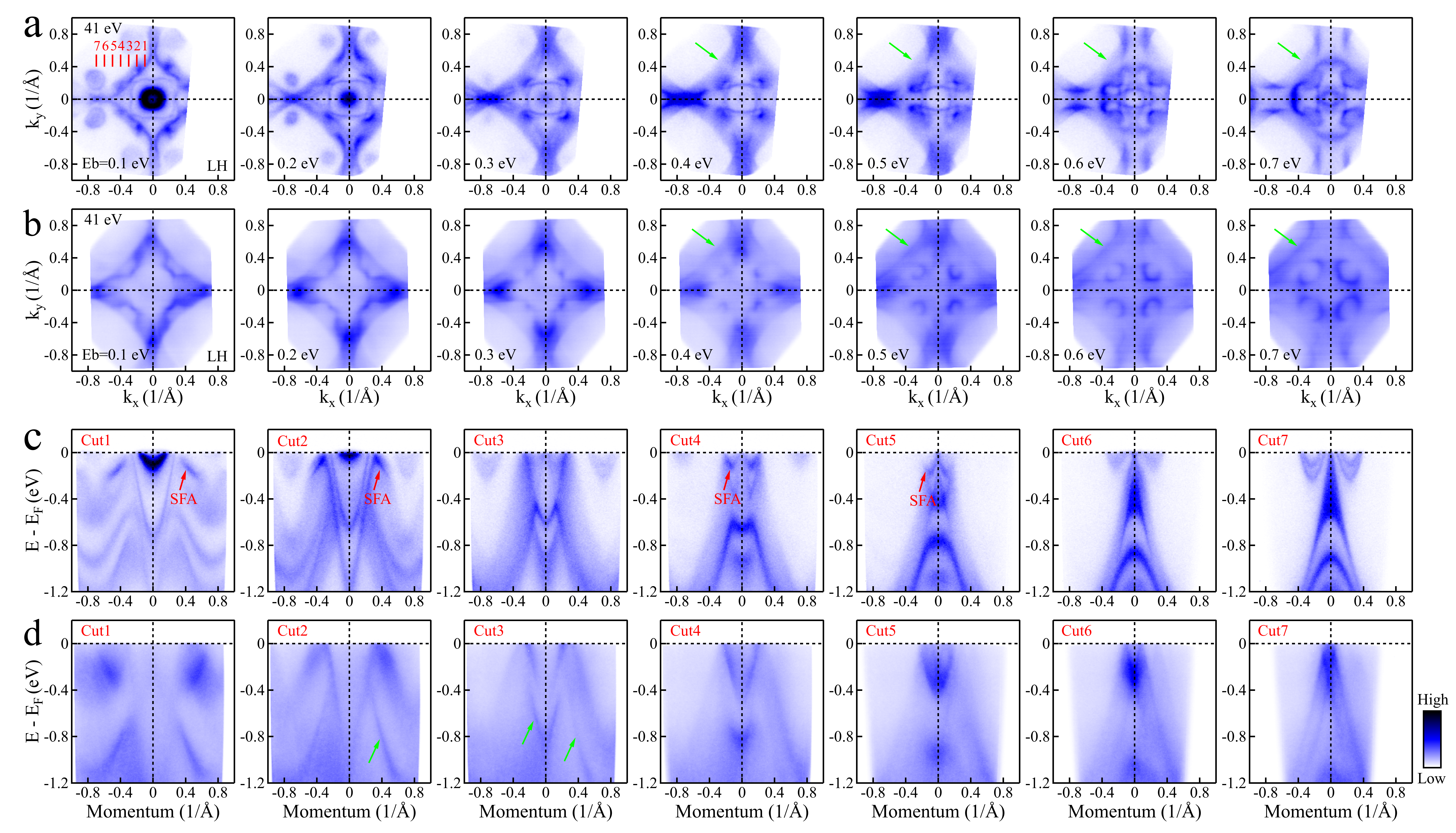}
\end{center}
\caption{\textbf{Comparison of constant energy contours and band structure of flat and uneven surfaces.} (a) The constant energy contous of NdAlSi measured in the area 1 (red circle in Fig. 2a) of sample with photon energy of 41 eV under LH polarization. (b) The constant energy contous of NdAlSi measured in the area 2 (orange circle in Fig. 2k) of sample with photon energy of 41 eV under LH polarization. (c-b) The band structure measured in the area 1 (c) and area 2 (d) of sample along Cut1-Cut7 directions with photon energy of 41 eV under LH polarization. The cut positions are marked by red lines in (a). The SFAs are marked by red arrows in (c).
}
\label{S1}
\end{figure*}

To compare the electronic structure measured on a flat and uneven surfaces in more detail, we carried out constant energy contour measurements, as shown in Fig.~\ref{S1}a-\ref{S1}b. Comparing the difference between them can help distinguish features of surface and bulk states. It can be found that although the electronic structure measured on the flat surface is dominated by the surface states, the bulk features can still be found at some energies. Fig.~\ref{S1}c-\ref{S1}d show the band structures measured along Cut1-Cut7 (red lines in Fig. ~\ref{S1}a) which covers half of the BZ. From the position of the Cuts in Fig.~\ref{S1}a, it can be seen that Cut1, Cut2, Cut4 and Cut5 pass right through or near the position of the surface Fermi arc (SFA). Therefore, the SFA projections can be seen in the band measurements of the corresponding Cut in Fig.~\ref{S1}c (marked by red arrows). Compared to the corresponding bands measured in area 2 (Fig.~\ref{S1}d), the SFAs completely disappear. In addition, the bulk states measured on the uneven surface do not show obvious degradation compared with the flat surface (Fig.~\ref{S1}c-\ref{S1}d), and some bands even look sharper (marked by green arrows in Fig.~\ref{S1}b and~\ref{S1}d).\\

{\bf 2. Comparison of band structures between fresh surface and 10 hours after cleavage}

\begin{figure*}[tbp]
\begin{center}
\includegraphics[width=1\columnwidth,angle=0]{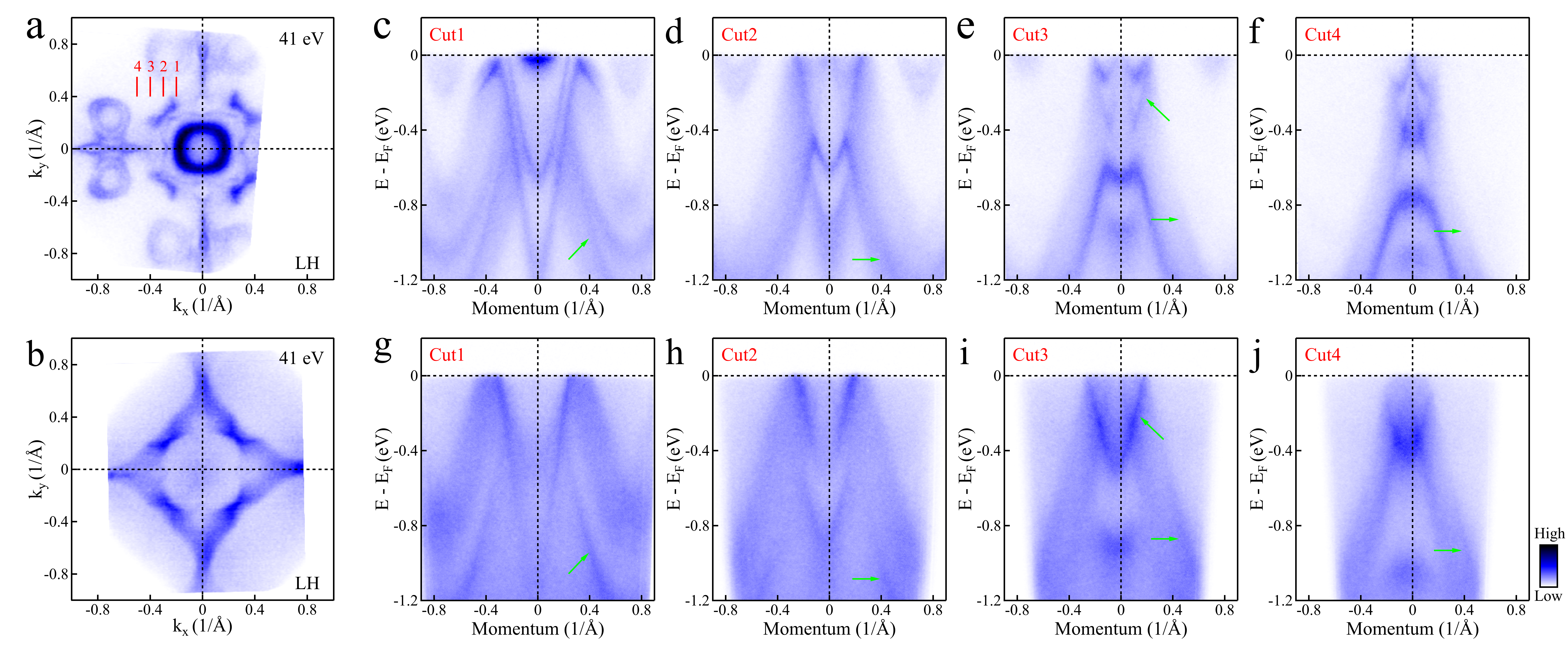}
\end{center}
\caption{\textbf{Comparison of band structure between fresh surface and 10 hours after cleavage.} (a) The constant energy contours of NdAlSi measured in the area 1 (red circle in Fig. 2a) of the fresh surface (a) and 10 hours after cleavage (b) with photon energy of 41 eV under LH polarization. (c-f) The band structures measured in the area 1 of fresh surface along Cut1-Cut4 directions with photon energy of 41 eV under LH polarization. (g-j) Similar measurements as (c-f) but measured 10 hours after cleavage.
}
\label{S2}
\end{figure*}

To compare the electronic structure measured on a fresh flat surface and 10 hours after cleavage in more detail, we did band structure measurements along Cut1-Cut4 (red lines in Fig.~\ref{S2}a), as shown in Fig.~\ref{S2}c-\ref{S2}j. The band structure measured on the flat surface 10 hours after cleavage look similar to that of the uneven fresh surface. Interestingly, after 10 hours of measurement, some of the bands looked sharper than when measured on the fresh surface (marked by green arrows in Fig.~\ref{S2}c-\ref{S2}j). Therefore, we conclude that after 10 hours, the surface of the sample has become disordered in a similar way as the fresh uneven surface. This disorder seems to affect only the surface electronic structure as some of the bulk bands become sharper. A potential explanation is that the impurities are adsorbed at low temperatures, which would remove charge and weaken the bonds between Nd atoms and the surface. This increased mobility of Nd atoms results in a disordered surface with irregular clusters, similar to an uneven fresh surface.

\vspace{3mm}

\newpage

%\begin{figure*}[tbp]
%\begin{center}
%\includegraphics[width=1\columnwidth,angle=0]{FigS3_EDCStacks.png}
%\end{center}
%\caption{\textbf{Energy distribution curves stacking of band structures.} (a) The same band structure measurement as Fig. 3k. (b) EDCs stacking of band structure in (a). (c) The same band structure measurement as Fig. 3o. The red line in (c) markes the band dispersion guided by eyes. (d) The EDCs stacking of band structure in (c). The red diamonds in (d) mark the peak positions of band guided by eyes. 
%}
%\label{S3}
%\end{figure*}

%\begin{figure*}[tbp]
%\begin{center}
%\includegraphics[width=1\columnwidth,angle=0]{FigS4_BandVsHvT.png}
%\end{center}
%\caption{\textbf{Photon energy dependent band structures.} (a-d) Photon energy dependent band structures measured on fresh surface of flat sample with photon energies of 30 eV (a), 41 eV (b), 49 eV (c) and 53 eV (d) along $\overline{X}-\Gamma-\overline{X}$ direction. (e-h) The similar measurements as (a-d) but measured on the surface 10 hours after the sample cleaved. (i-l) The similar measurements as (a-d) but measured on the surface 20 hours after the sample cleaved. (m) The integrated EDC of the time dependent bands measured with photon energy of 30 eV which is normalized at energy range of -1.2 eV to -1.0 eV. (n-p) The similar measurements as (m) but measured with photon energies of 41 eV (n), 49 eV (o) and 53 eV (p).
%}
%\label{S4}
%\end{figure*}